\documentclass[epj]{svjour}
\usepackage{graphics}
\usepackage{amssymb}
\begin{document}

\title{Constraints on the time-scale of nuclear breakup from thermal hard-photon emission}
\author{R. Ortega\inst{1,2}\thanks{\emph{Present address: Physikalisches Institut, Universit\"{a}t Heidelberg, Philosophenweg 12 D-69120 Heidelberg, Germany}} 
\and D. d'Enterria\inst{1,2,3}\thanks{\emph{Present address: Nevis Laboratories, Columbia University  538 W. 120th St. NY 10027, USA}} 
\and G. Mart\'{\i}nez\inst{2,3}  \and D. Baiborodin\inst{4}  \and H. Delagrange\inst{2,3} \and J. D\'{\i}az\inst{5}\and
F. Fern\'andez\inst{1} \and \\ H. L\"{o}hner\inst{6} \and T. Matulewicz\inst{7}  \and R.W. Ostendorf\inst{6} \and S. Schadmand\inst{8} \and Y. Schutz\inst{2,3}\thanks{\emph{Present 
address: CERN, CH-1211 Gen\`eve, Switzerland}}  \and P. Tlusty\inst{4} \and R. Turrisi\inst{3}\thanks{\emph{Present address: INFN-Padova, Via Marzolo 8, 
35131 Padova, Italy}}\and V. Wagner\inst{4} \and H.W. Wilschut\inst{6} \and N. Yahlali\inst{5}}

\institute{Grup de F\'{\i}sica de les Radiacions, Universitat Aut\` onoma de Barcelona 08193, Catalonia \and SUBATECH, 4, rue Alfred Kastler BP20722
44307 Nantes Cedex 3, France \and GANIL, Grand Acc\'el\'erateur National d'Ions Lourds, IN2P3-CNRS, DSM-CEA, 14076 Caen Cedex 5,  France
\and Institute of Nuclear Physics, 20568 \~{R}e\~{z}, Czech Republic \and Institut de F\'{\i}sica Corpuscular, Universitat de Valencia-CSIC, Dr Moliner 50,
46100 Burjassot, Spain \and Kernfysisch Versneller Instituut, 9747 AA Groningen, The Netherlands \and Institut of Experimental Physics, Warsaw University,
00681 Warsaw, Poland \and Institut f\"ur Kernphysik, Forschungszentrum J\"ulich, D-52457 J\"ulich, Germany}
\date{Received: date / Revised version: date}
\abstract{Measured  hard photon multiplicities from second-chance
  nucleon-nucleon collisions are used in combination with a kinetic thermal model, to estimate the break-up 
times of excited nuclear systems produced in nucleus-nucleus reactions at intermediate energies. The obtained nuclear break-up time for the $^{129}${Xe} + $^{nat}${Sn}
reaction at 50{\it A} MeV is $\Delta$$\tau$ $\approx$  100 -- 300 fm/$c$ for all reaction centralities. 
The lifetime of the radiating sources produced in seven other different heavy-ion reactions studied by the TAPS experiment are
consistent with $\Delta$$\tau$  $\approx$ 100 fm/$c$, such relatively long thermal photon emission times  do not support the interpretation of nuclear 
breakup as due to a fast spinodal process for the heavy nuclear systems studied.
} 
\maketitle
\section{Introduction}
\label{intro}
Nucleus-nucleus collisions are the only available tool to explore in
the laboratory the different domains of the nuclear phase diagram. In
heavy-ion (HI) collisions
at intermediate energies (bombarding energies between 20{\it A} MeV and 100{\it A} MeV) the projectile and target traverse each other in a time scale 
$t$ $\lesssim$ 50 fm/$c$ forming a compressed and excited transient system that usually disassembles into several intermediate-mass-fragments (IMF) in a 
process known as "nuclear multifragmentation" \cite{reports}. The exact physical mechanism that drives this breakup is up to date an open issue. Two opposite
scenarios have been considered: a fast breakup from an expanding source, consistent with a spontaneous spinodal mechanism, and a sequential slower 
breakup from a thermally equilibrated source. The knowledge of the relative importance of these two mechanisms is essential to find-out  a possible 
connection between nuclear multifragmentation and the occurrence of a ``liquid-gas"-like phase transition \cite{Viola} in nuclear systems with excitation 
energies  in the range $\epsilon$$^{\star}$\,=\,(3 -- 8){\it A} MeV, where a flattening of the ``caloric curve", signaling a possible liquid-gas phase transition, has
been observed \cite{Poch95,nato02c,nato02,trautmann2004}. 
To discern between both possible mechanisms it is necessary to establish the breakup time of the multifragmenting system. 
% Experimentally, following the trace of the chronology of the reaction is 
%a very ambitious objective, mainly at the
%intermediate stages of the reaction which we are interested in, given
%the dynamic evolution of the experimental \emph{finite} system. 

The experimental observable used in the present analysis in order to probe the thermodynamical state of the  produced transient nuclear systems,
and in particular, to delimit the nuclear breakup time is thermal hard
photon emission. At intermediate-energy  HI collisions, photons 
above 30 MeV (hard photons) issue from bremsstrahlung in incoherent  proton-neutron collisions \cite{Nife90}. The main hard photon
contribution, the so-called ``direct" component,  is emitted from first-chance nucleon-nucleon ($NN$) collisions in the preequilibrium stage
of the reaction. A second and softer flash of hard photons is emitted from secondary $NN$ collisions in the later stages \cite{gines,yves}
from a thermalizing source \cite{dav01}. Hence, the experimental hard photon spectrum measured in HI reactions has
been found to be  well reproduced by the sum of two exponential distributions \cite{yves}:
\begin{equation}
 \frac{d\sigma}{dE_\gamma}\,=\,K_d\:e^{-E_\gamma/E_0^d}+K_t\:e^{-E_\gamma/E_0^t}
\label{eq:2exponentials}
\end{equation}
where {\it d} and {\it t} stand for direct and thermal, respectively, and the factors $K_{d,t}$ are related to the intensity of
each source. 
%\\The power of thermal bremsstrahlung \emph{photons} as experimental tool relies on the conjunction of two facts:
%they are emitted from an {\it equilibrated} source, and they are ``clean" probes undistorted by final-state interactions
%with the surrounding nuclear medium. Using a kinetic thermal model \cite{ther}, the thermal slope has been recently found 
%to be an efficient direct thermometer to provide a direct measurement
%of the temperature of the excited source \cite{dav02}.
The power of first chance bremsstrahlung photons is the fact that they result
from incoherent collisions. This allowed one to equate the observed
intensity and slope of the gamma spectrum with the number  and hardness of collisions respectively. With the
kinematic thermal model \cite{ther} this information can also be
obtained from secondary collisions; the soft spectrum now corresponds
to the number of collisions integrated over the lifetime of the
equilibrated source and its formation, and the slope parameter is related
to the average temperature of the source \cite{dav02}. Although
multiple collisions lead to bremsstrahlung quenching (see {\it e.g.}
\cite{goeth}) it is believed not to play an important role because we
exploit the photon spectrum above $E_{{\gamma}cut}$=30 MeV. To
observe about 10\% quenching the collision rate of the nucleons in the
thermal source should not exceed $E_{{\gamma}cut}$/hbar=1.5 $\cdot$
10$^{-1}$ (fm/c)$^{-1}$, in the model, without considering Pauli
blocking, the rate typically is 1.3 $\cdot$ 10$^{-1}$ (fm/c)$^{-1}$.
Thus the time scales derived in this work  are not limited by the
quenching phenomena.

The analysis presented in this paper has two different goals. First, we want to elucidate the dependence of the thermal photon temperature
on the reaction centrality in $^{129}${Xe} + $^{nat}${Sn}  reactions at 50{\it A} MeV; at this energy, this heavy and symmetric system provides 
good conditions for observing the effects of a possible nuclear liquid-gas phase transition, and moreover, the charged-particle distributions and
related observables have been studied in detail for this reaction  by the INDRA Collaboration (see e.g.\cite{Mari97,Mari98,Gou00,Bell02,Steck01} and 
references therein). The second objective consists in assessing the nuclear breakup time-scales using the measured second-chance photon multiplicities
and a thermal model that describes well the observed spectra. 
This investigation is  performed for 7 different reactions studied by the TAPS Collaboration: $^{86}${Kr} + $^{58}${Ni} at 60{\it A} MeV \cite{gines}, 
$^{181}${Ta} + $^{197}${Au} at 40{\it A} MeV \cite{gines},  $^{208}${Pb} + $^{197}${Au} at 30{\it A} MeV \cite{gines}, $^{36}${Ar} + $^{107}${Ag} at 60{\it A} MeV \cite{dav01}, $^{36}${Ar} + $^{197}${Au} 
at 60{\it A} MeV \cite{dav01},  $^{36}${Ar} + $^{58}${Ni} at 60{\it A} MeV \cite{dav01} and  $^{129}${Xe} + $^{nat}${Sn} at 50{\it A} MeV.
The paper is organized as follows: in Sect. \ref{sec:experiment} the experimental setup  is described. In Sect. \ref{sec:ident} the
particle identification methods are presented. The results obtained from the analysis of the inclusive hard photon spectrum are summarized 
in Sect. \ref{sec:inclusive}. In Sect. \ref{sec:excl} we focus on the ${Xe}$ + $^{nat}${Sn} exclusive measurements; first, we present the
experimental centrality classes as well as the method employed to estimate the average impact parameter; next, we present the direct
and thermal hard photon measurements for each centrality class.  In  Sect. \ref{sec:thermo}, the thermodynamical properties 
extracted from the thermal hard photon signal are presented. In particular in \ref{subsec:model} the results on the lifetimes of the thermal hard photon sources
obtained  for the different reactions are presented. Section \ref{sec:summ} gives a summary.

\section{Experimental setup}
\label{sec:experiment}
The experiment E-300 was performed at the ``Grand \\Acc\'el\'erateur National d'Ions Lourds" (GANIL) located in Caen, France, during 13
days of total beam time in May 1998. The GANIL accelerator system  delivered  a Xe$^{46+}$ beam of 50{\it A} MeV with a bunch rate of 10 MHz
and a nominal intensity  of 5 nA, around 68 ions per pulse. The Xe ions impinged  on a $^{\rm nat}$Sn target of 1 mg/cm$^2$  thickness 
(with associated interaction probability of 1.9$\cdot10^{-3}$ per burst).
\begin{figure*}[htbp]
\begin{center}
%\resizebox{0.55\textwidth}{!}{\includegraphics{figures/xesn_ganil.eps}}
\resizebox{0.55\textwidth}{!}{\includegraphics{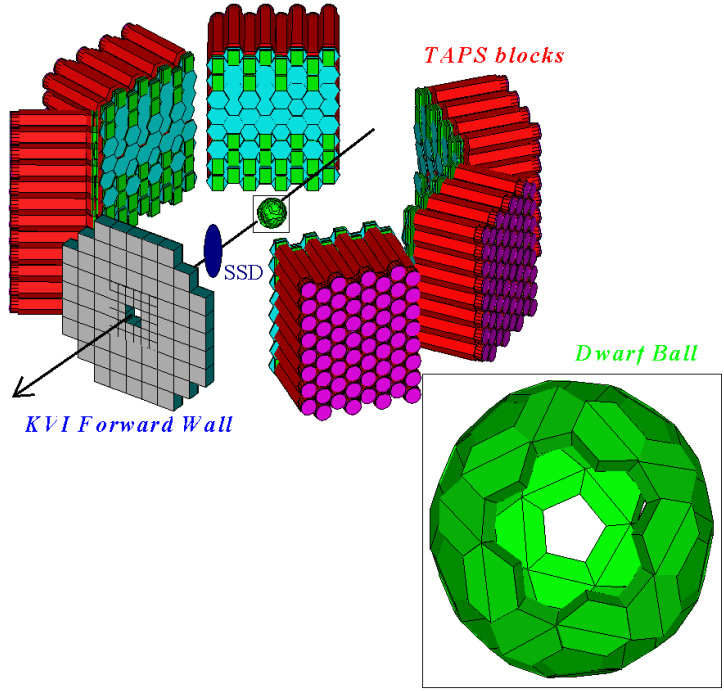}}
\end{center}
 %\vspace*{5cm}       % Give the correct figure height in cm
\caption{\it General layout of the experimental setup used to study the reaction $^{129}${Xe}+$^{nat}${Sn} at 50{\it A} MeV at GANIL. 
The detector system  consists of the Silicon Strip Detector (SSD), the Washington University Dwarf Ball (DB), the TAPS electromagnetic
calorimeter and the KVI Forward Wall (FW).}
\label{fig:det.eps}
\end{figure*}
The experimental setup  consisted of the "Two Arms Photon Spectrometer" TAPS associated with three particle multidetectors:
the GANIL Silicon Strip Detector (SSD), the Washington University ``Dwarf Ball" (DB) and the KVI ``Forward Wall" (FW). A general layout
of the experimental setup is presented in Fig. \ref{fig:det.eps}.
 
The TAPS  electromagnetic calorimeter \cite{novot} consists of 384
scintillation modules. Each one consisted of a BaF$_2$ hexagonal
crystal, 2.95 cm inner radius and 25 cm length (corresponding to 12$X_0$ radiation lengths) \cite{lav,gab}, and a Charged Particle Veto (CPV) detector
of  plastic (NE102A) scintillator. TAPS offers a  large solid angle coverage 
allowing  for high quality angular distribution measurements  in a  wide  energy range, from statistical (E$_\gamma$$>5$ MeV) to the hardest
photons (E$_\gamma \approx$ 200 MeV). In this experiment TAPS modules were arranged in six blocks of 64 modules (8 $\times$ 8), which were 
placed in an almost symmetrical configuration in the horizontal plane around the target, and at an average distance of 56 cm,
covering  about 20\% of the full solid angle.

The GANIL  silicon strip detector telescope SSD  was placed at 19 cm
downstream  the target inside the vacuum chamber, covering the angular
range of $2.2^\circ \leq \theta \leq 10.3^\circ$. This device, consisting of a first disk 150 $\mu$m thick with 64 semicircular strips and 
of a second disk 500 $\mu$m thick with 128 radial strips,  was sensitive to projectile-like fragments (PLF) emitted in binary reactions and 
intermediate-mass fragments (IMF) emitted in the forward direction from more dissipative reactions. 

The Washington University ``Dwarf Ball" (DB) \cite{Stra90} covered  an angular range of $31^\circ \leq \theta \leq 168^\circ$ around 
the target. This detector system allows to identify light-charged particles (LCP) and IMF. The Dwarf Ball (DB)   consists of 64
phoswich detectors forming a sphere with an inner radius of 41.5 mm. Each phoswich detector is made of a thin plastic scintillator and of
a CsI(Tl) inorganic crystal scintillator. The plastic scintillator type (Bicron BC400 or BC446) depends  on the covered angle and so does
its thickness, 10 $\mu$m and 40 $\mu$m  for the most backward and forward modules, respectively. 
The CsI(Tl) crystal glued on the back of each plastic is 4 mm to 8 mm thick, depending also on the angle.

The KVI ``Forward Wall"(FW) \cite{Luke93}, consisted of 92 plastic scintillator phoswiches,   detects and identifies LCP and IMF 
(up to Z$\approx$15) emitted in the forward direction. In our experimental setup, this detector was placed downstream  74 cm
away from the target covering the forward hemisphere behind the SSD, $2.5^\circ \leq \theta \leq 25^\circ$ and the whole azimuthal  range. 
Each FW phoswich is composed of two organic plastic scintillators which are heat-pressed together; a 1 mm thick ``fast" ($\tau$\,=\,2.4 ns)
NE102A plastic followed by a 50 mm thick ``slow" ($\tau$\,=\,320 ns) NE115 plastic.

This complete experimental setup allowed correlated measurements of photons and charged-particles,  necessary to study the dependence
of the hard-photon production on the reaction topology. The photon
data were recorded with the minimum bias trigger for neutral particles
 ($\gamma$*DB), 
which signaled events with at least one neutral hit in TAPS of energy $E_{\gamma}>$ 10 MeV  detected in coincidence with one or more 
charged particles  in the Dwarf Ball.  The minimum bias (MB) reaction
trigger was defined by the condition of detecting at least a charged
particle in the Dwarf Ball.
\section{Particle identification}
\label{sec:ident}
\begin{figure}
\begin{center}
\resizebox{0.52\textwidth}{!}{\includegraphics{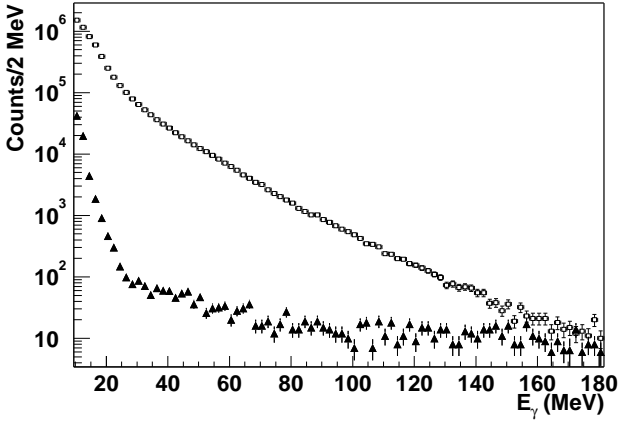}}
\end{center}
\caption{\it Inclusive raw photon spectrum measured in the center-of-mass frame for the reaction $^{129}${Xe} + $^{nat}${Sn} at 50{\it A} 
MeV. The triangles show the cosmic ray background contribution.}
\label{fig:en_raw.eps}
\end{figure}

Photons are identified in TAPS by means of  a Pulse Shape (PSA) vs. time-of-flight (TOF) analysis and the CPV information.
Photons and leptons induce a stronger intensity of the fast BaF$_2$ light component compared to hadronic particles, producing therefore a 
higher ratio of the fast to the slow energy components in the BaF$_2$ (PSA). Photons and  relativistic electrons are located in time within
a prompt peak centered at TOF=1.87 ns (with a resolution $\sigma$ = 340 ps \cite{Orte03}); these particles produce hence a recognizable contour in
the PSA-TOF spectrum. In order to separate photons from relativistic electrons the information (fired or not) delivered by the CPV of each
module is used. Nevertheless, given that cosmic muons have mainly vertical trajectories not firing  the CPV, and have random TOF, they can
be misidentified as photons. This is avoided by defining  a PSA-TOF cosmic contour several nanoseconds far from the photon contour but with
the same dimensions. Subsequently, the energy and angular spectra  measured within this contour (see Fig. \ref{fig:en_raw.eps}) are 
subtracted from the raw photon energy and angular spectra respectively. \\The $\Delta$E\,-\,E telescope technique \cite{Orte03,got} allows
the charge identification of  LCP and IMF detected in the DB and in the FW, as well as  of the PLF and IMF impinging on  the SSD. The light
measured in the first stage of each multidetector (the first component of each phoswich in the  DB and the FW, or the first disk in the SSD)
and the signal of the second component are plotted in bidimensional histograms in which the different curves correspond to different Z. 
In the DB,  additional LCP isotopic identification was also possible exploiting the pulse shape properties of the CsI(Tl)  light.
\section{Inclusive hard photon spectra in  $^{129}${Xe} + $^{nat}${Sn}  at 50{\it A} MeV reactions}
\label{sec:inclusive}
Figure \ref{fig:en_raw.eps}  shows the inclusive raw photon energy spectrum measured in the $^{129}${Xe} + $^{nat}${Sn} reaction from the 
collection of $5.8\cdot10^6$ photons with $E_\gamma>10$ MeV detected in TAPS and transformed into the CM frame ($\beta_{AA} \approx
\beta_{NN} \approx$ 0.15). After cosmic-ray subtraction, and for energies $E_\gamma>30$ MeV this spectrum is well described by the 
double source fit of Eq. \ref{eq:2exponentials} (Fig. 3). The direct component exhibits an inverse slope parameter $E_0^d$\,=\,(15.6 $\pm$ 1.0) MeV
whereas the thermal slope parameter is $E_0^t$\,=\,(7.0 $\pm$ 0.6) MeV. Both slopes are in agreement with the direct \cite{pol} and thermal 
\cite{dav02} photon slope systematics measured in intermediate $A+A$
reactions. Thermal hard photons account for 22\% of the 
total yield above 30 MeV. The strength 
and intensity of the thermal emission are also confirmed  by a double source analysis of the hard photon angular distribution; however,
due to the symmetry of the system, the NN and AA source velocities cannot be disentangled \cite{raquel02}. 
 
The direct (thermal) hard photon multiplicity, i.e. the number of direct (thermal) hard photons emitted per nuclear reaction, has been  
obtained via the following expression:
\begin{equation}
\label{eq:photon_mult}
%M_{\gamma}^{d,t}\,=\,\frac{\sigma^{d,t}_{\gamma}}{\sigma_R}\,=\,\frac{\varepsilon_R}{\varepsilon_{\gamma}}\cdot
%\left(\frac{SD}{T}\right)_{\gamma*DB} \cdot \left(\frac{SD}{T}\right)_{DBor} \cdot
%\frac{N_{\gamma}^{d,t}}{N_{R}}
M_{\gamma}^{i}\,=\,\frac{\sigma^{i}_{\gamma}}{\sigma_R} 
\end{equation}
where $i$ stands for direct or thermal, $\sigma^{i}_{\gamma}$ is the experimental hard photon cross section determined by integrating 
the corresponding  $d\sigma/dE_{\gamma}$ spectrum above $E_{\gamma}$ = 30 MeV, and $\sigma_R$ = (5300  $\pm$ 600) mb is the experimental 
total reaction cross section for Xe+Sn, obtained from the charged particle distribution measured in the DB with the MB reaction
trigger. We report the  direct and thermal hard photon cross sections and multiplicities in Table \ref{tab:cross_exp}.
 \begin{table}
\caption{\it Summary of the experimental   inclusive hard photon ($E_\gamma >$ 30 MeV) results obtained for the 
$^{129}${Xe}\,+\,$^{nat}${Sn} reaction at 50{\it A} MeV. For the direct and thermal $\gamma$ components we quote:
inverse slope $E_0$, relative intensity $I$, source velocity $\beta$, cross-section $\sigma_{\gamma}$, multiplicity $M_{\gamma}$, 
and (only for the direct component)  hard photon probability $P_{\gamma}$.}
\label{tab:cross_exp}
\begin{center}
\vspace{5mm}
\begin{tabular}{lll}
\hline\noalign{\smallskip} 
&  Direct  & Thermal \\ 
\noalign{\smallskip}\hline\noalign{\smallskip}
$E_0$(MeV) &  15.6 $\pm$ 1.0 &  7.0  $\pm$ 0.6\\
$I$ (\%)&  78 $\pm$ 1 &  22 $\pm$ 1\\
$\beta$&  0.15 $\pm$ 0.01 &  0.16 $\pm$ 0.01\\
$\sigma_{\gamma}$(mb)  & 4.9 $\pm$ 0.6 &  1.4  $\pm$ 0.2\\
$M_{\gamma}$ &  (9.3 $\pm$ 0.8)$\cdot$ $10^{-4}$ & (2.6 $\pm$ 0.3)$\cdot$ $10^{-4}$ \\
$P_{\gamma}$ & (7.8 $\pm$ 0.7) $\cdot$ $10^{-5}$ &    -\\
\noalign{\smallskip}\hline
\end{tabular}
\end{center}
\end{table}
The direct hard photon probability $P_{\gamma}^d$, i.e. the probability to produce a hard photon in a first-chance proton-neutron 
collision, is determined from the experimental direct hard photon multiplicity:
 \begin{equation}
\label{eq:P_pn}
P_{\gamma}^{d}\,=\,\frac{\sigma_{\gamma}^d}{\sigma_{R}\cdot \langle N_{pn}\rangle_b}\,=\,\frac{M_\gamma^{d}}{\langle N_{pn}\rangle_b}
\end{equation}
where $\langle N_{pn}\rangle_b$ is the number of first chance pn collisions averaged over  impact parameter, calculated from the geometrical
``equal-participant" model of  Nifenecker and Bondorf \cite{Nife85}. $\langle N_{pn}\rangle_b$ is = 11.79  for the Xe on Sn system. The measured
$P_{\gamma}^d$ in $^{129}${Xe}\,+\,$^{nat}${Sn} reactions at 50{\it A} MeV is (7.8 $\pm$ 0.7) $\cdot$ $10^{-5}$. We note that we do not quote
a thermal hard photon probability, since we should determine the average number of secondary (and not primary) pn collisions, from which the
thermal component originates, and this number is not well determined experimentally nor theoretically. More details of the inclusive  analysis 
can be found in \cite{Orte03}. 
%This first thermal hard-photon
%measurement was followed by the question of the evolution of the
%thermal emission with the reaction centrality. In the
%following subsections this analysis is presented.
\begin{figure}
\begin{center}
\resizebox{0.53\textwidth}{!}{\includegraphics{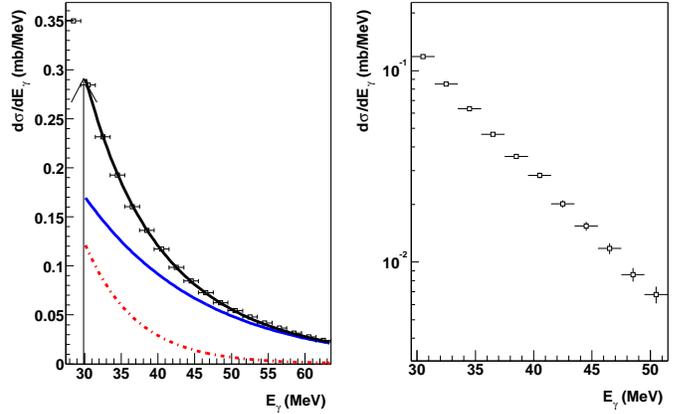}}
\end{center}
\caption{\it  Left panel: experimental inclusive hard photon spectrum measured for the reaction $^{129}${Xe} + $^{nat}${Sn} at 50{\it A}  in the
range E$_\gamma$\,=\,30\, -\,65 MeV, fitted to the
Eq. \ref{eq:2exponentials}, the solid (dashed) line is the direct
(thermal) exponential distribution. Right panel: inclusive thermal hard
photon spectrum obtained after subtracting  the fitted direct
component.} 
\label{fig:expo}
\end{figure} 
\section{Exclusive hard photon analysis in $^{129}${Xe} + $^{nat}${Sn}  at 50{\it A} MeV reactions}
\label{sec:excl}
\subsection{Selection of centrality classes}
\label{sec:Reaction_classes}
Six centrality classes of $^{129}${Xe} + $^{nat}${Sn} collisions have been selected to investigate the
dependence of the thermal hard photon production on the reaction centrality. This choice results from a compromise between requiring a 
wide impact parameter coverage and obtaining enough hard photon statistics  for each centrality class. 
\begin{table*}
\begin{center}
\caption{\it Conditions on the fragment multiplicity in the SSD ($M^{SSD}$), the  charged particle multiplicity (IMF + LCP) in the
DB ($M^{DB}$) and the FW ($M^{FW}$) that define each centrality class. The contribution to the reaction cross-section,  and to the nuclear
reactions detected by the  minimum bias photon trigger ($\gamma$*DB) for each centrality class are  reported.}
% The condition on the charged particle multiplicity for the inclusive measurements is also shown.}
\label{tab:exclusive1}
%\begin{center}
\begin{tabular}{llll}
\hline\noalign{\smallskip} 
Centrality class   & Multiplicity condition & \%$\sigma_R$   & \%$\sigma_R(\gamma$*DB)\\ 
\noalign{\smallskip}\hline\noalign{\smallskip}
   A or Periph. & $M^{DB+FW}$ = 1-2 and $M^{SSD}$  ($\theta < 5$) = 0  & 57\% & 23\%\\
   B &  $M^{DB+FW}$  = 1-2 & 67\% &  26\%\\
   C &  2 $<M^{DB+FW}<8$ & 38\% & 58\%\\
   D & 5 $<M^{DB+FW}<9$ & 10\% & 24\%\\
   E & 6 $<M^{DB+FW}<10$ & 7\% & 18\%\\
   F or Central &  9 $<M^{DB+FW}<15$ & 2\% & 6\% \\
 %  Incl. & $M^{DB}$$\geq$ 1 & & \\
\noalign{\smallskip}\hline
\end{tabular}  
\end{center}
\end{table*}
We report in Table \ref{tab:exclusive1} the conditions on the fragment multiplicities measured in the SSD, Dwarf Ball and
Forward Wall that define the selected centrality bins.
\begin{table*}
\caption{\it Characteristics of the six centrality classes considered 
for Xe+Sn reactions at 50A MeV. For each centrality bin we quote: average charged particle multiplicity
$M_{cp}$ and  impact parameter $\langle$b$\rangle$$_{geo}$
estimated with the geometrical model of Cavata et al. \cite{Cava90}; direct photon multiplicity
($M^{d}_{\gamma}$), average number of first chance proton-neutron
collisions $\langle$$N_{pn}$$\rangle$ obtained as described in \cite{Rie,Mart},  and associated
averaged impact parameter $\langle$b$\rangle$$_{\gamma}$ estimated using the "equal-participant" model \cite{Nife85}.} 
\label{tab:exclusive2}
\begin{center}
\begin{tabular}{llllll}  
\hline\noalign{\smallskip}
Centrality class & $\langle$$M_{cp}$$\rangle$ &$\langle$b$\rangle$$_{geo}$ (fm)& $M^{d}_{\gamma}$  &
$\langle$$N_{pn}$$\rangle$  & $\langle$b$\rangle$$_{\gamma}$(fm)\\ 
\noalign{\smallskip}\hline\noalign{\smallskip}
   A  or Periph. & 1.5&9.4 $\pm$ 0.8 & (3.2 $\pm$ 0.2)$\cdot10^{-4}$ &  4.1 $\pm$ 0.6 & 9.2 $\pm$ 0.3\\
   B   & 1.5& 9.4 $\pm$ 0.8 & (4.1 $\pm$ 0.3)$\cdot10^{-4}$ & 5.2 $\pm$ 0.6 & 8.8 $\pm$ 0.3\\
   C &  4.7 &  5.2 $\pm$ 0.4  &  (1.5 $\pm$ 0.1)$\cdot10^{-3}$ & 18.6  $\pm$ 2.2 & 5.9 $\pm$ 0.5 \\
   D   &  6.8&3.3 $\pm$ 0.3& (2.4 $\pm$ 0.2)$\cdot10^{-3}$ & 30.7 $\pm$ 3.7 &  4.0 $\pm$ 0.6\\
   E   & 7.8& 2.6 $\pm$ 0.2&  (2.8 $\pm$ 0.2)$\cdot10^{-3}$ & 36.1 $\pm$ 4.4&  3.1 $\pm$ 0.7\\
   F or Central  &11.4&  0.9 $\pm$ 0.1 &  (3.8 $\pm$ 0.4)$\cdot10^{-3}$ & 48.9  $\pm$ 6.7  & 1.3 $\pm$ 1.1 \\
\noalign{\smallskip}\hline
\end{tabular}
\end{center}
\end{table*}
\subsection {Estimation of the impact parameter for minimum bias data}
\label{subsec:parameter}
In order to estimate the impact parameter b  in $^{129}${Xe} +
$^{nat}${Sn}  all reactions, recorded by means of the MB trigger, we  apply the geometrical method proposed by 
Cavata {\it et al.} \cite{Cava90}, which relies on the monotonous decrease of  $M_{CP}$ as a function of b.
An alternative method to estimate the impact parameter in experiments where  hard photon emission can be analyzed  is based on the 
correlation of the direct hard $\gamma$ yield with the impact parameter 
\cite{Rie,Mart}. In this method, for a given reaction class, the experimental average number of first chance $pn$ collisions 
$\langle$$N_{pn}$$\rangle$ is estimated, making use of Eq. \ref{eq:P_pn}, from the  experimentally measured   $M_\gamma^{d}$
and from  the $P_{\gamma}^{d}$, which is considered to depend only on the Coulomb corrected bombarding energy. The obtained 
$\langle N_{pn} \rangle$ values are related to the impact parameter b
by  the geometrical ''equal-participant"  model of of  Nifenecker and Bondorf \cite{Nife85}.
We note  that if  the impact parameter is obtained directly from the total hard photon yield, as commonly done, instead of only from the 
direct component, the deduced impact parameter might be distorted for reactions where a thermal photon  contribution cannot be neglected.
The  average impact parameter for each reaction class, obtained by means of the direct hard photon multiplicities and by means of the 
charged particle multiplicities measured in the DB,  are reported in Table \ref{tab:exclusive2}. We find a good agreement between both 
methods, and in addition the whole impact parameter range is covered.
%Despite the agreement between
%$\langle$b$\rangle$$_{\gamma}$ and
%$\langle$b$\rangle$$_{geo}$, the
%   monotonous
%   variation of $M_{cp}$ with $b$ assumed by the geometrical method is not valid for central
%   collisions, since the maximum $M_{cp}$ does not correspond
%   to a real
%   $b$=0.
%  Concerning centrality class A,  since for this centrality class the
%  condition on the total charged particle multiplicity is
%%   the same as for reaction B ($M^{DB+FW}$ =
%  1-2), the average
%   impact parameter  cannot be
%  distinguished from the one of reaction B from the geometrical
%  method.
\subsection {Determination of the impact parameter for reactions where a hard photon is produced}
\label{subsec:b_shift}
We want to exploit the properties of the thermal hard photon component as a tool to extract the thermodynamical state of the radiating
source. It is therefore of crucial importance to determine as accurately as possible any  bias introduced in the estimation of the impact
parameter for  particular reactions where a hard photon is produced. 

\begin{figure}[htbp]
 \begin{center}
 \resizebox{0.5\textwidth}{!}{\includegraphics{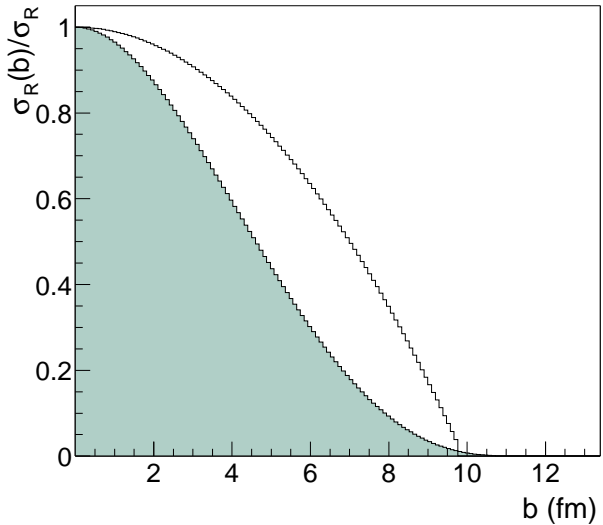}}
 \end{center}
 \caption{\it Evolution of $\sigma_R$(b)/$\sigma_R$   as a function
 of the impact parameter without (empty histogram) and with (filled histogram)
 the condition that a hard photon is produced.}
 \label{fig:shift}
 \end{figure}

Within a given centrality class, defined by a range in the $M_{cp}$ distribution,  hard photons are mostly emitted in  reactions with 
high  $M_{cp}$ values, characterized by a large projectile-target overlap, and consequently by a high number of $pn$ collisions. 
% Hence, the $\langle$b$\rangle$ calculated by means of the previous method(s)  may be overestimated
 %for reactions where a hard photon is produced. In these reactions, the production probability is proportional to 
%$\sigma^{\gamma}_{R}(b)$. 
This bias through lower impact parameters can be  reproduced
by  the geometrical ''equal-participant" model; the reaction cross-section as a function of b for reactions where a hard photon is produced,
$\sigma^{\gamma}_{R}$(b), is modulated  by the product of $\sigma_{R}$(b) and  $N_{pn}$(b). As a consequence, the ratio between the  reaction
cross-section $\sigma_R^{\gamma}$(b) and the total (integrated over b) $\sigma_R^{\gamma}$ is a representable function sensitive to the 
cross section deviation through lower b (see Fig. \ref{fig:shift}).
It is  thus possible to estimate the average impact parameter for reactions where a hard photon is produced by means of an extension of 
the geometrical model of Cavata {\it et al.}, correlating the experimental $\sigma^{\gamma}_R$($M_{cp}$)/$\sigma^{\gamma}_R$ and the 
theoretical $\sigma^{\gamma}_R(b,N_{pn})/\sigma_R$ obtained from the geometrical ''equal-participant" model.
The average b of each centrality class has been obtained with this method after weighting  the b($M_{cp}$) by the contribution of each
$M_{cp}$ within the centrality class. Table \ref{tab:impact_shift} reports the $\langle$b$\rangle$ estimated for each RC for reactions 
where a hard $\gamma$ is emitted.
 \begin{table}[htbp]
 \caption{\it Estimated average impact parameter $\langle$b$\rangle$
 for each of the six centrality classes considered.}
 \label{tab:impact_shift}
%\begin{center}
\begin{tabular}{ll}  
\hline\noalign{\smallskip}
Centrality class &$\langle$b$\rangle$$_{geo}$ (fm)    \\ 
\noalign{\smallskip}\hline\noalign{\smallskip}
  Class A or Peripheral & 7.6 $\pm$ 0.3\\
  Class B & 7.2 $\pm$ 0.3  \\
  Class C &  4.2 $\pm$ 0.4  \\
  Class D  &2.8 $\pm$ 0.4 \\
  Class E  & 2.2 $\pm$ 0.6 \\
  Class F or Central &  1.0 $\pm$ 0.9  \\
\noalign{\smallskip}\hline
\end{tabular}
%\end{center}
\end{table}
\subsection {Centrality dependence of the  hard photon production}
\label{sec:centrality} 
\begin{figure}[htbp]
 \begin{center}
 \resizebox{0.51\textwidth}{!}{\includegraphics{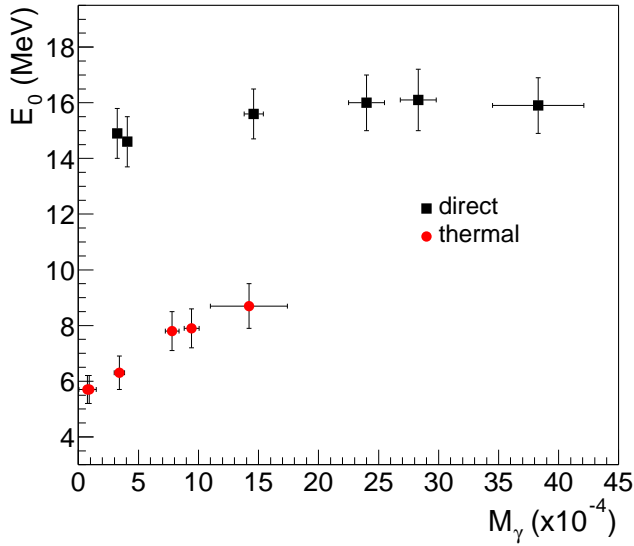}}
 \end{center}
 \caption{\it Direct and thermal hard photon slopes vs their respective multiplicities for the six Xe+Sn centrality classes considered.}
 \label{fig:slope_multi.eps}
 \end{figure}

The two-exponential
fit of Eq. \ref{eq:2exponentials} reproduces the hard photon energy
spectrum of each one of the six centrality classes, from peripheral to
central events. The corresponding direct and thermal slopes parameters and intensities  are listed in Table \ref{tab:exclusive3}. Whereas 
the measured $E^t_0$(b) differ by  up to 24\% from the inclusive  value, the direct slope variation amounts to  less than 5\% of the 
inclusive $E^d_0$.
\begin{table*}[htbp]
\caption{\it Characteristics of the hard photon spectra measured
in the six different centrality classes: direct ($E_0^{d}$) and
thermal ($E_0^{t}$) slopes, ratio of thermal to total
hard-photon intensities ($I_{t}/I_{tot}$), and  thermal
photon multiplicity ($M^{t}_{\gamma}$).}
\label{tab:exclusive3}
\begin{center}
\begin{tabular}{llllll}  
\hline\noalign{\smallskip}
Centrality class & $E_0^{d}$ (MeV) & $E_0^{t}$ (MeV) &
$I_{t}/I_{tot}$(\%) & $M^{t}_{\gamma}$ & $\chi^2/ndf$   \\ 
\noalign{\smallskip}\hline\noalign{\smallskip}
A or Periph. & 14.9 $\pm$ 0.9 & 5.7 $\pm$ 0.5 & 19 $\pm$ 1 &    (7.6 $\pm$ 0.7)$\cdot10^{-5}$ &  1.5   \\
B & 14.6 $\pm$ 0.9 & 5.7 $\pm$ 0.5 & 19 $\pm$ 1 &  (8.9 $\pm$ 1.1)$\cdot10^{-5}$ & 1.2\\
C & 15.6 $\pm$ 0.9 & 6.3 $\pm$ 0.6 & 19 $\pm$ 1   &   (3.4 $\pm$ 0.4)$\cdot10^{-4}$ & 1.0\\
D  & 16.0 $\pm$ 1.0 & 7.8 $\pm$ 0.7 & 25 $\pm$ 1&  (7.8 $\pm$ 0.6)$\cdot10^{-4}$ & 1.4\\
E & 16.1 $\pm$ 1.1 & 7.9 $\pm$ 0.7 & 25 $\pm$ 1 &  (9.4 $\pm$ 0.6)$\cdot10^{-4}$ &  1.2  \\
F  or Central & 15.9 $\pm$ 1.0 & 8.7 $\pm$ 0.8 & 27 $\pm$ 2 &  (1.4 $^{+0.2}_{-0.3}$)$\cdot10^{-3}$ &  1.0  \\
\noalign{\smallskip}\hline
\end{tabular}
\end{center}
\end{table*}
%The aforesaid preequilibrium causes would be the ultimate responsible  of the variation of the thermal slope.
The measurement of an almost constant  direct slope confirms that the $E_0^d$  is   an observable that depends just on the  bombarding
energy \cite{yves}. Only for very peripheral reactions it is expected to observe a decrease of the direct photon slope mainly due to the 
decrease of the Fermi momentum in the surface of the  nucleus \cite{Mart}.
%This excitation energy dependence is connected to the former preequilibrium phase space conditions.
In Fig. \ref{fig:slope_multi.eps} the values of the thermal (direct)  slopes are displayed as a function of the  thermal (direct) 
multiplicities. The  direct hard photon slope is independent of the direct hard photon multiplicity. This shows that the increase of the
direct hard $\gamma$ multiplicity with the centrality is just a consequence of the associated increase of the number of first-chance
pn collisions. On the other hand, the thermal slope exhibits a linear dependence on the thermal multiplicity. This trend, observed for the 
first time, indicates that, assuming that the size of the thermalized source remains almost constant with 
centrality, the thermal production is  sensitive to the excitation  of
the nuclear system.
This result is supported by the inclusive thermal systematics collected by the TAPS Collaboration: the thermal slopes and thermal 
multiplicities divided by the size of the system, scale  with the energy available in the nucleus-nucleus center-of-mass \cite{dav01,Orte03}. 
% Therefore,
%the  linear relation $M_{\gamma}^t$ and $E_{0}^t$ indicates that
%the evolution of the thermal hard photon yield with the centrality
%is only  due to  the increase of the excitation energy and, thus,
%of the energy available in secondary collisions.
 \subsection{Exclusive analysis of the  GDR emission}
 \label{subsec:gdr}

Exploiting the wide dynamic range of TAPS, we have estimated the $\gamma$ yield
from the Giant Dipole Resonance (GDR) decay emitted in peripheral $^{129}${Xe} + $^{nat}${Sn}  reactions. The main motivation of this 
investigation was  to   identify a possible thermal bremsstrahlung and GDR mixture in the low energy range ($E_{\gamma}\approx$ 30 MeV) of the hard photon
spectra, which could distort the measurement of the thermal hard photon yield. Peripheral reactions exhibit the
highest probability of GDR$\gamma$ contamination because the reached excitation energies  can still be not high enough for suppression of the collective mechanisms responsible
of the GDR$\gamma$ emission \cite{Suo96}, and because of the softer observed thermal emission. So the GDR analysis in
peripheral reactions allowed us to conclude about a possible GDR$\gamma$ contamination for all
reaction classes.

In order to perform the GDR$\gamma$ analysis, the direct and thermal hard photon components have been subtracted from the total  photon 
spectra. This subtraction is done by extrapolating the double hard $\gamma$ exponential fit down to the lowest energy, given by the TAPS 
LED threshold of 10 MeV. This  extrapolation technique is routinely applied in GDR measurements from HI reactions 
(see {\it e.g.} \cite{Gaar87,LeF94,Suo96}). 
 %However in those measurements the hard photon distribution is parameterized by an only exponential.

\begin{figure}[htbp]
 \begin{center}
\resizebox{0.53\textwidth}{!}{\includegraphics{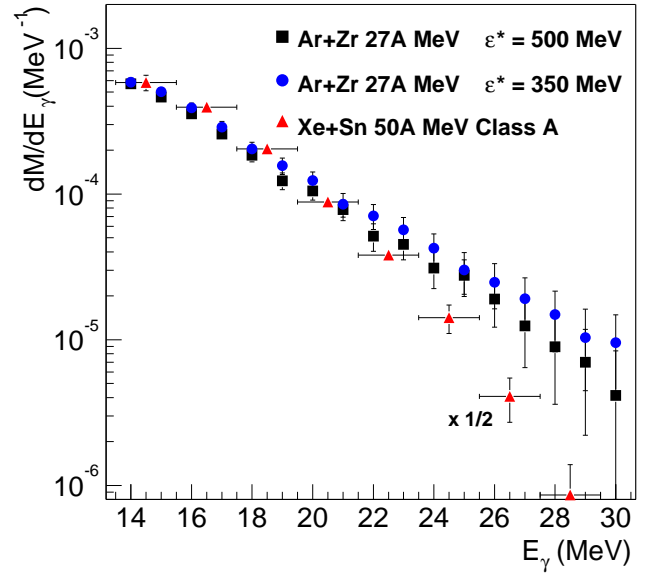}}
 \end{center}
\caption{\it Spectra in the GDR region for the $^{36}${Ar} + $^{90}${Zr} reaction \cite{Suo96},
for two different excitation energies: $\epsilon$$^{\star}$\,=\,350 MeV (dots),
$\epsilon$$^{\star}$\,=\,500 MeV (squares)\cite{Suo96}, compared 
to the experimental $\gamma$ spectrum obtained in peripheral $^{129}${Xe} + $^{nat}${Sn}
reactions (triangles) after subtraction of the hard-photon yield 
(extrapolated down to 10 MeV).}
\label{fig:gdr.eps}
\end{figure}

We have compared the GDR spectra measured for  peripheral centrality classes A and B with the  GDR distributions measured with the
multidetector  MEDEA and  reported by T. Suomij\"{a}rvi {\it et al.} \cite{Suo96}. In this work, the GDR$\gamma$ emission from  hot nuclei
of A $\approx$ 115 formed in the $^{36}${Ar} + $^{90}${Zr} at 27{\it A} MeV reaction is investigated in detail. The comparison of the 
GDR$\gamma$ yield measured in this analysis and the one measured for peripheral reactions in our work comes easily to mind when considering
the similarities in mass and in  excitation energy of the emitting sources:

\begin{itemize}
\item{In peripheral $^{129}${Xe} + $^{nat}${Sn} reactions  the quasi-projectile and quasi-target
   are close to the $^{129}${Xe} and $^{nat}${Sn} masses. Assuming that both, quasi-target and quasi-projectile, develop 
collective GDR oscillations after the collision, the photon spectrum 
measured in the peripheral centralities scaled by a factor of $\sim$1/2 
can be compared to the GDR $\gamma$ yield emitted by a nuclei of A $\sim$  115;}
\item{In spite of the different geometry and bombarding energy of both reactions, the excitation energies reached
  at large impact parameters in $^{129}${Xe} + $^{nat}${Sn} reactions
 ($\sim$ 3{\it A} MeV, see next section) are comparable with the  $\epsilon$$^{\star}$ attained in
 the $^{36}${Ar} + $^{90}${Zr}  MeV
 reaction at 27{\it A}. Besides,  the GDR yield measured  for the $^{36}${Ar} +
$^{90}${Zr} system has been found to remain constant with the
excitation energy \cite{Suo96} (see Fig. \ref{fig:gdr.eps}).}
%This saturation is in good agreement with the GDR quenching \cite{Gaar87} observed at high excitation energies
%$\epsilonn^{\star} >$ 3 MeV.}
\end{itemize}
The GDR yields (the $\gamma$ spectra after subtraction of the bremsstrahlung contribution) for the $^{36}${Ar} + $^{90}${Zr} reaction for 
$\epsilon$$^{\star}$\,=\,350 MeV and for $\epsilon$$^{\star}$\,=\,500 MeV from \cite{Suo96}, and  the GDR yield
estimated for the $^{129}${Xe} + $^{nat}${Sn} centrality class A\footnote{the GDR yield measured for centrality classes A and B are almost 
identical, so only the yield for class A is shown in  the figure.} are plotted together in Fig. \ref{fig:gdr.eps}. The global  compatibility 
of the three distributions assures the validity of the estimation of the GDR yield for the peripheral Xe+Sn hard photon spectra. The difference
found in the 25-30 MeV region  between the TAPS and MEDEA measurements is due to the fact that in the latter the subtracted hard photon distribution
is parameterized by a single exponential. Taking into account  that for peripheral reactions the total photon multiplicity  is around  9
$\cdot 10^{-5}$ at $E_{\gamma}$ = 30 MeV and that the GDR yield at this energy is found to be lower than  $10^{-5}$ (see Fig. \ref{fig:gdr.eps}), we conclude 
that the GDR contamination in the low energy range of the hard photon spectra  can be neglected. This result  is in good agreement with
the highest energy GDR gamma contribution predicted from the systematics. Indeed, taking the centroid energy of the GDR as $E_{GDR}\,=\, 76.5A^{-1/3}$ 
and the maximum width $\Gamma_{GDR}\approx$ 12 MeV \cite{Suo96,Berm75}, the GDR emission neither  from the Xe projectile nor from the Sn target
are expected to exceed  28 MeV.
\section{Derived (thermo)dynamical properties}
\label{sec:thermo}
\subsection{Nuclear Temperature}
Although, due to their long mean free path in nuclear matter, thermal hard photons are not in thermal equilibrium with the source, their spectral 
shape is directly correlated to the temperature of  the radiating nuclear fragment. As in our previous  analyses \cite{dav02}, we use the kinetic
thermal model of Neuhauser and Koonin \cite{ther} to calculate the relation between the inverse slope $E_0^t$ and the temperature $T$ of the nuclear 
system. The electromagnetic radiation rate  emitted by a hot and equilibrated nuclear fragment is expressed in this model as a function of the local
density and temperature of the source. It is assumed that bremsstrahlung in proton-neutron  collisions is the main source of thermal photon emission
for $E_{\gamma}$\,$>$\,30 MeV. The hard photon spectra thus calculated  are accurately approximated by an exponential with slope $E_0^t$ in agreement
with the experimental data \cite{dav02}. The evolution of  $E_0^t$
values extracted from a fit above 30 MeV with the temperature, T, is quantitatively 
well described by the following linear expression:
 \begin{equation}
T(\mbox{MeV})\,=\,(0.78\,\pm\,0.02)\cdot E_0^t(\mbox{MeV})
\label{eq:T_E0t}
\end{equation}
in the region T $\approx$ 3\,-\,10 MeV and  $\rho$ $\approx$ (0.3\,-\,1.2)$\rho$$_0$ \cite{dav02}.
We have applied this photon  thermometer to extract the temperature of the radiating system produced in  the different $^{129}${Xe} + $^{nat}${Sn} 
centralities studied in this work. The obtained temperatures are reported in Table \ref{tab:tempe}.
\begin{table}[htbp]
\caption{\it Nuclear temperatures $T$ estimated through Eq.
\ref{eq:T_E0t} from the experimental thermal hard photon
slopes  $E_0^{t}$ of the inclusive  and exclusive hard photon
spectra measured in  $^{129}${Xe} + $^{nat}${Sn} at 50{\it A}
MeV.}
\label{tab:tempe}
\begin{center}
\begin{tabular}{lll}  
\hline\noalign{\smallskip}
Centrality class & $E_0^{t}$ (MeV) & T (MeV)  \\ 
\noalign{\smallskip}\hline\noalign{\smallskip}
  Inclusive & 7.0 $\pm$ 0.6  &  5.5  $\pm$ 0.8\\
   A or Peripheral & 5.7 $\pm$ 0.5  &  4.4  $\pm$ 0.4\\
   B & 5.7 $\pm$ 0.5  & 4.4  $\pm$ 0.4 \\
   C &  6.3 $\pm$ 0.6  & 4.9  $\pm$ 0.5 \\
   D  & 7.8 $\pm$ 0.7 & 6.1  $\pm$ 0.5\\
   E  &  7.9 $\pm$ 0.7 & 6.2  $\pm$ 0.5  \\
  F or Central &  8.7 $\pm$ 0.8   & 6.8 $\pm$ 0.6 \\
\noalign{\smallskip}\hline  
\end{tabular}
\end{center}
\end{table}
The temperature of the nuclear system for central collisions (centrality class F, $b/b_{max}$ $\leq$ 0.1) is a 50\% $\pm$ 20\% higher than for  peripheral
collisions (centrality class A, $b/b_{max}$ $\approx$ 0.6). 

\subsection{Lifetime of the thermalized system}
\label{subsec:model} The multiplicity of hard photons emitted by an
equilibrated nuclear fragment  can be calculated by means of the model of Neuhauser and Koonin from the following expression \cite{dav02}:

\begin{equation}
 M_{\gamma}^{NK}=\,\,\int\,d^{3}x\,\int\,dt\,\int_{30}^{\infty}\,\frac{dR_{\gamma}^{NK}(T,\rho)}{dE_{\gamma}}\,dE_{\gamma}
 \label{eq:mulko}
 \end{equation}
where  $R_{\gamma}^{NK}$($T$,$\rho$) is the  rate of photons emitted  from a thermal nuclear fragment with temperature $T$ and density $\rho$. 
From this model, $R_{\gamma}^{NK}$($T$,$\rho$)  is found to   scale with $\sim$\,$T^{6.7}$ and to be proportional to $\rho$. The quadratic dependence of
the photon rate on $\rho$  expected naively  turns out to be linear due to the Pauli blocking. The photon rate can be hence expressed as 
$R_{\gamma}^{NK}$$\approx$\,$R_{0}^{NK}$$\cdot$ $T^{6.7}\cdot \rho$, with the constant $R_{0}^{NK}=1.1\cdot 10^{-13}$ MeV$^{-6.7}\cdot$(fm/$c$)$^{-1}$. 
We can  further simplify Eq. \ref{eq:mulko}, by assuming that the temperature is uniform in the volume $V$ and roughly  constant during the
emission time,
\begin{equation}
M_{\gamma}^{NK} \approx\,V \cdot \Delta\tau \cdot R_{0}^{NK} \cdot
T^{6.7}\cdot \rho
 \label{eq:mulko_2}
\end{equation}
where $V$ and $\Delta\tau$ are respectively the volume and the lifetime of the radiating nuclear source. The volume can be approximated as the
ratio of the sum and the projectile and target nucleons~\footnote{We neglect here emission of particles during preequilibrium.} 
%which e.g. for the Xe+Sn reaction  accounts for 15-25\%  of the system mass [ref?]
over the nuclear density,\\ $V$$\approx$\,(A$_t$\,+\,A$_p$)/$\rho$. 

\begin{figure}[htbp]
\begin{center}
\resizebox{0.53\textwidth}{!}{\includegraphics{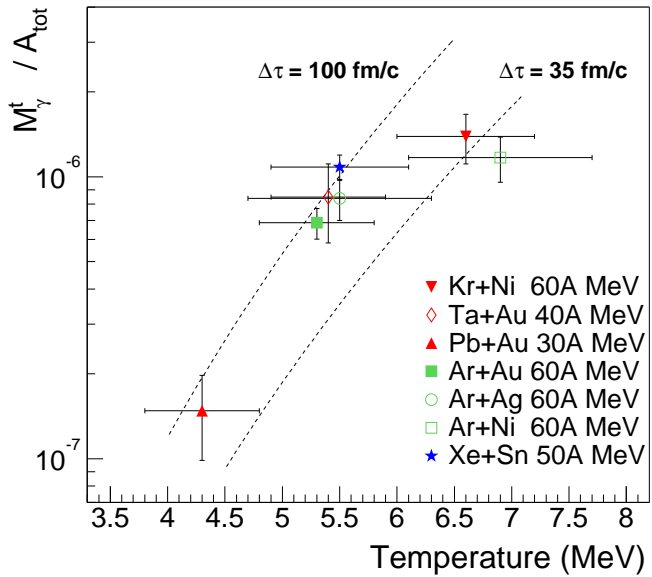}}
\end{center}
\caption{\it Experimental thermal bremsstrahlung multiplicity, $M_\gamma^{t}$, divided by
the  size of the colliding system, $A_{tot}$\,=\,$A_t$+$A_p$, plotted as a
function of the extracted nuclear temperatures, for the different
reactions studied by the TAPS Collaboration. The lines correspond
to two different values of the source lifetime estimated with Eq.
\ref{eq:mulko_3}.} 
\label{fig:temps1.eps}
\end{figure}

With these  simplifying assumptions, the thermal hard photon yield, scaled to the relative size of the system, can be written  as:
\begin{equation}
\frac{M_{\gamma}^{NK}}{(A_t\,+\,A_p)} \approx\,\Delta\tau \cdot
R_{0}^{NK}\cdot T^{6.7}
 \label{eq:mulko_3}
\end{equation}
This relation, which is independent of the density, can be exploited to estimate the lifetime $\Delta\tau$ of the thermal equilibrated source, by 
fitting  Eq. (7) to the experimentally measured multiplicities (reported in Table 5) as a function of the temperature.  Figure \ref{fig:temps1.eps}
shows the experimental (inclusive) $M_\gamma^{t}$, divided by  size of the system $A_{tot}$\,=\,$A_t$+$A_p$, as a function of the estimated $T$,
for the Xe+Sn and the six other reactions studied \cite{dav02}. The lines correspond to the calculated $M_{\gamma}^{NK}$ as a function of the $T$ 
for two different lifetimes ($\Delta\tau$=100 fm/$c$ and $\Delta\tau$=35 fm/$c$).

\begin{figure}[htbp]
 \begin{center}
 \resizebox{0.54\textwidth}{!}{\includegraphics{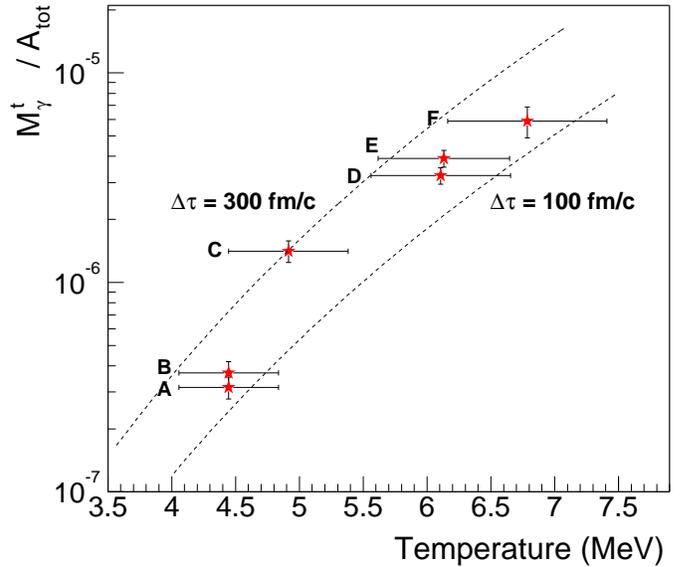}}
\end{center}
\caption{\it Thermal hard photon multiplicities, $M_\gamma^{t}$,  divided by the total size
of the system, $A_{tot}$\,=\,$A_t$+$A_p$, plotted as a function of the
nuclear temperature extracted from  Eq. \ref{eq:T_E0t}, for
the different $^{129}${Xe} + $^{nat}${Sn} centrality classes.The
lines correspond to two different  values of source lifetime computed with
Eq. \ref{eq:mulko_3}.} \label{fig:temps2.eps}
\end{figure}

The calculated multiplicities successfully reproduce the  experimental thermal hard photon yields measured for all the reactions (see Figs.
\ref{fig:temps1.eps} and \ref{fig:temps2.eps}), with $\Delta\tau$ values  about 100 fm/$c$, which are consistent with the expected lifetime of an 
equilibrated source (see e.g. \cite{wada04}).
This result confirms the validity of the thermal model, since it reproduces both experimental observables, the slope of the thermal hard photon spectrum
and the thermal multiplicity, with the temperature as the single input parameter. In Figs. \ref{fig:temps1.eps} and \ref{fig:temps2.eps}, it can be observed
that all the systems studied have a lifetime $\Delta\tau$ of the order or larger than 100 fm/$c$, except the two lighter systems at 60{\it A} MeV,
$^{86}$Kr+$^{58}$Ni and $^{36}$Ar+$^{58}$Ni, which seem to survive a shorter  $\Delta\tau$ $\sim$ 35 fm/$c$, similar to the transit time of the colliding ions. 
In these two reactions, the achieved excitation energy is also higher than in the rest of
reactions \cite{dav02}. This difference in $\Delta\tau$ might indicate that equilibrated nuclear sources
 break faster  when they have a smaller size or higher excitation energies, as observed in 
 \cite{wada04,Beau00,He01}.
%Long times are associated with low excitation energies and short
%times with larger values \cite{Beau00}.
The experimental thermal hard photon multiplicities measured for the $^{129}${Xe} + $^{nat}${Sn} centrality classes exhibit a trend compatible 
with lifetimes of the order of  100-300 fm/$c$ (Fig.\ref{fig:temps2.eps}). In our measurements, within the data uncertainty, we do not observe any
change of the lifetime of the equilibrated fragment with the impact parameter of the collision for Xe+Sn. The relatively large lifetime values obtained,
jointly with the measurement of an important production of thermal hard photons, suggest that, at least for the investigated Xe+Sn reactions, the fireball
does not  undergo an instantaneous breakup from a rapidly expanding
state (spinodal decomposition). 

\subsection{Caloric curve constructed from  Xe\,+\,Sn exclusive measurements}
\label{subsec:caloric_curve}
As  done in a previous paper for inclusive thermal photon measurements \cite{dav02}, we have measured the relationship between the estimated temperature
and the excitation energy, $\epsilon^\star$, of the hot equilibrated fragment formed in  Xe\,+\,Sn reactions at the different centralities studied in this
work. At variance with experiments that measure the excitation energy  with charged particle detectors, we cannot determine ``calorimetrically" 
$\epsilon^{\star}$ from our data. The excitation energy of  Xe quasiprojectiles produced from peripheral to central collisions has been deduced
by Steckmeyer {\it et al.} \cite{Steck01} as a function of the charged particle multiplicity measured by the INDRA detector. The methods described in
\cite{Steck01} are based on the determination of the velocity of the quasiprojectile.  The excitation energy is calculated from the kinetic energies of 
products belonging to the source. We have employed the geometric method of Cavata {\it et al.} \cite{Cava90} to relate the average impact parameter 
of each  Xe\,+\,Sn reaction class  with the corresponding value of $M_{cp}^{\mbox{\tiny{\it{INDRA}}}}$, in order to obtain the measured
value of   $\epsilon$$^{\star}$  for each centrality class. The
estimated $M_{cp}^{\mbox{\tiny{\it{INDRA}}}}$ and $\epsilon$$^{\star}$ values  are reported in Table \ref{tab:energyIndra}. 

\begin{table}[htbp]
\caption{\it Values of the charged particle multiplicity measured
with INDRA, $M_{cp}^{\mbox{\tiny{\it{INDRA}}}}$,  and of the estimated excitation energy,  $\epsilon^\star$,
for each Xe+Sn centrality class with averaged impact parameter
$\langle$b$\rangle$.}
  \label{tab:energyIndra}
\begin{center}
\begin{tabular}{llll}  
\hline\noalign{\smallskip}
Centrality class & $\langle$b$\rangle$ (fm) & $M^{INDRA}_{cp}$ &
$\epsilon$$^{\star}$({\it A} MeV)  \\ 
\noalign{\smallskip}\hline\noalign{\smallskip}
 Inclusive & 3.8 $\pm$ 1.0 & 25  $\pm$ 1 & 5.6 $\pm$ 1.0 \\
   A or Peripheral & 7.6 $\pm$ 0.3 & 14  $\pm$ 1 & 2.8 $\pm$ 0.3 \\
   B & 7.2 $\pm$ 0.3 & 16  $\pm$ 1 & 3.3 $\pm$ 0.4 \\
   C &    4.2 $\pm$ 0.4 & 24  $\pm$ 1 & 5.2 $\pm$ 0.6 \\
   D  &  2.8 $\pm$ 0.4 & 27  $\pm$ 1 & 6.1 $\pm$ 0.7\\
   E  &  2.2 $\pm$ 0.6 & 29    $^{+2}_{-1}$  & 6.7 $\pm$ 0.7  \\
  F or Central &  1.0 $\pm$ 0.9 & 32 $^{+5}_{-2}$ & 7.7  $^{+1.4}_{-1.0}$(8.7 $\pm$ 1.5) \\
\noalign{\smallskip}\hline
\end{tabular}
\end{center}
\end{table}

\begin{figure}[htbp]
\begin{center}
\resizebox{0.5\textwidth}{!}{\includegraphics{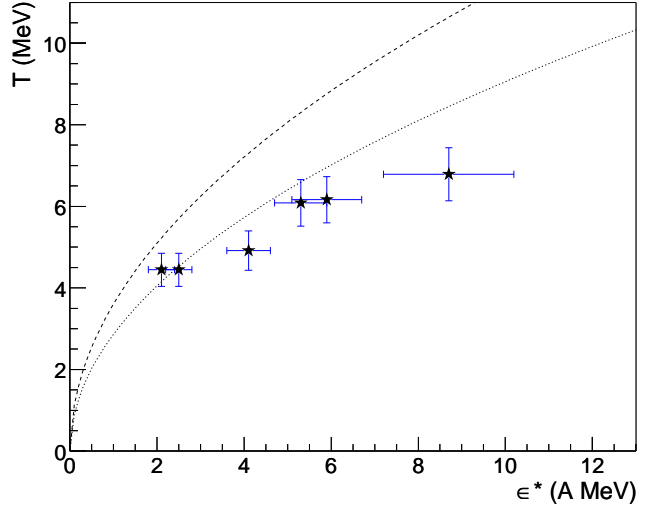}}%caloric_excl2.eps %caloric_excl3.eps only XE+Sn
\end{center}
\caption{\it Caloric curve constructed from the thermal hard
photon slope measurements for the $^{129}${Xe} + $^{nat}${Sn}
reaction at 50{\it A} MeV (Table 6)  and the estimated excitation energies 
(Table 7), compared to the Fermi liquid model curves with inverse level-density parameters
K=8/A MeV (dashed line) and K=13/A MeV (dotted line).}
\label{fig:caloric2.eps}
\end{figure}

The errors of $\epsilon$$^{\star}$ take into account the uncertainty of $\langle$b$\rangle$ and a  $\sim$10\% systematic error due to the reconstruction
of the quasiprojectile \cite{Steck01}. As expected, the excitation energy decreases linearly with increasing impact parameter. Since the highest measured charged
particle multiplicity does not correspond to a real $\langle$b$\rangle=0$, the geometrical method used to relate $\langle$b$\rangle$ and 
$M_{cp}^{\mbox{\tiny{\it{INDRA}}}}$ is not accurate enough for  central collisions. We therefore also quote in parentheses for centrality class F  the value of 
$\epsilon^\star$ predicted for  central collisions by the model of Natowitz {\it et al.} \cite{nato86}.
The caloric curve obtained is displayed in
Fig. \ref{fig:caloric2.eps}. The temperature increases smoothly with
the excitation energy, the ($\epsilon^{\star}$,T) pairs obtained are
below the expected Fermi fluid curves ($T = \sqrt{K\epsilon^\star}$
(MeV) with K = 8/A -- 13/A MeV covering the known range of average
nuclear level density parameters~\cite{Nerlo-Pomorska:2005}). 
Peripheral reactions (classes A and B)
lead to excitation energies $\epsilon^{\star}$$\approx$ 3{\it A} MeV,
near those identified as the onset of the leveling region of the 
caloric curve for nuclear systems of size
A= 180--241 \cite{nato02c}.

%By  varying the centrality of the collision, instead of using
%Bseveral bombarding energies or systems, the obtained caloric curve
%Bshould be more self consistent, since the extracted temperatures
%Bhave no uncertainties attribuTable to different experimental
%Bconditions or analysis procedures. However, aside the limited
%B$\epsilon^{\star}$ range covered in a single reaction system and the
%Bdifficulties to determine $\epsilon^{\star}$(b), a possible
%Bdrawback, mostly in the case of very peripheral collisions, is the
%Buncertainty in the mass of the equilibrated composite system.
\section{Summary}
\label{sec:summ}
Hard photon ($E_\gamma >$ 30 MeV) emission has been studied  in $^{129}${Xe} + $^{nat}${Sn} at 50{\it A} MeV  for six bins of impact-parameter. 
The thermal component has been determined from a double source analysis of the measured spectra. For each centrality class, the temperature of the 
produced nuclear systems has been extracted from the measured thermal hard photon slope, by applying a thermal nuclear bremsstrahlung model  that 
successfully reproduces the measured thermal hard photon spectral shape and multiplicities. The temperature obtained exhibits a small but systematic
increase with decreasing impact parameter, from $T$ = 4.4 MeV for the most peripheral collisions ($b/b_{max}$ $\approx$ 0.6) to $T$ = 6.8 MeV   
for the most central  class ( $b/b_{max}$ $\leq$ 0.1). The amount of second-chance bremsstrahlung emission measured in each centrality bin is well 
described by the thermal model by assuming emission times of the order $\tau\approx$ 100 -- 300 fm/$c$ for all centralities. 
 %We have not observed here any change in the lifetime with the violence of the collision. 
The lifetimes of the produced equilibrated sources have also been estimated for six other heavy-ion reactions for which thermal bremsstrahlung emission has been 
measured by the TAPS Collaboration. The inclusive thermal hard photon multiplicities measured as a function of the temperature $T$ for the different
reactions suggest lifetimes $\tau$ $\sim$ 100 fm/$c$, i.e. between two and three times longer than the transit time of the colliding ions at these energies. 
Such relatively large time scale disfavor scenarios of instantaneous spinodal break-up and is more consistent with sequential fragmentation from a thermally
equilibrated source. It is worth noting, however, that for the two
smallest target-projectile combinations,  $^{86}${Kr}+$^{58}${Ni} and
$^{36}${Ar}+$^{58}${Ni} at 60A MeV, which are the systems with more energy available in the nucleus-nucleus center-of-mass, the experimental thermal photon
multiplicities seem to be better reproduced with  shorter lived equilibrated sources ($\tau \sim$ 35 fm/$c$).
%However, and given also the experimental uncertainties, the estimated lifetimes are larger than the expected for spinodal decomposition.

Finally, we present for the first time a  caloric curve, $\epsilon^\star(T)$, obtained from thermal hard photon measurements for different impact parameter bins of the same reaction,
Xe + Sn. Such a curve falls below the one expected for a Fermi liquid, in  agreement with the curve constructed from thermal hard photon slopes measured
in different heavy-ion reactions \cite{dav02}, and consistent also with the available systematics of nuclear caloric curves \cite{nato02c}.

\section{Acknowledgements}
 
We thank the GANIL accelerator staff for providing a high quality beam and technical support during the experiment. We thank J.C. Steckmeyer and the INDRA
Collaboration for providing the charged-particle multiplicity data, and the MEDEA Collaboration for providing the GDR data. This work has been in part 
supported by IN2P3-CICYT PN97-1 agreement, by the Dutch Foundation FOM, by the European Union HCM network under Contract No. HRXCT94066, and by DGICYT 
under contract FPA2000-2041-C02-01. R. Ortega aknowledges support
given by the Alexander von Humboldt Foundation. D.d'E. acknowledges support by the European Union TMR Programme (Marie-Curie Fellowship No. HPMF-CT-1999-00311).
 
% BibTeX users please use
% \bibliographystyle{}
% \bibliography{}

\end{document}